\documentclass[aps,preprint,nofootinbib,floatfix]{revtex4-1}
\usepackage{graphicx,color}
\usepackage[latin1]{inputenc}
\usepackage{amsmath,amssymb}
\usepackage{hyperref}
\usepackage{epstopdf}
\usepackage{geometry}                
\usepackage[normalem]{ulem}
\def\be{\begin{equation}}
\def\ee{\end{equation}}
\def\bea{\begin{eqnarray}}
\def\eea{\end{eqnarray}}
\newcommand{\gev}{~{\rm GeV}}
\newcommand{\tev}{~{\rm TeV}}
\newcommand{\mev}{~{\rm MeV}}

\begin{document}
\preprint{OSU-HEP-15-06}
\title{The search for mirror quarks at the LHC}

\author{Shreyashi Chakdar$^{1,2}$}
\email{chakdar@virginia.edu}
\author{K. Ghosh$^{1}$}
\email{kirti.gh@gmail.com}
\author{V. Hoang$^2$}
\email{vvh9ux@virginia.edu}
\author{P. Q. Hung$^{2,3}$}
\email{pqh@virginia.edu}
\author{S. Nandi$^{1}$}
\email{s.nandi@okstate.edu}

\affiliation{$^1$Department of Physics and Oklahoma Center for High Energy Physics,
Oklahoma State University, Stillwater, OK 74078-3072, USA \\
$^2$Department of Physics, University of Virginia, Charlottesville, VA 22904-4714, USA\\
$^3$Center for Theoretical and Computational Physics, Hue University College of Education, Hue, Vietnam.
}

\date{\today}                                           

\begin{abstract}

Observation of non-zero neutrino masses at a scale $\sim 10^{-1} - 10^{-2}$ eV  is a major problem in the otherwise highly successful Standard Model. The most elegant mechanism to explain such tiny neutrino masses is the seesaw mechanism  with right handed neutrinos. However, the required  seesaw scale is so high,  $\sim 10^{14}$ GeV, it will not have any collider implications. Recently, an explicit model has been constructed  to realize the seesaw mechanism with the right handed neutrinos at the electroweak scale. The model has a mirror symmetry having both the left and right  lepton and quark doublets  and singlets  for the same $SU(2)_W $ gauge symmetry. Additional Higgs multiplets have been introduced to realize this scenario. It turns out that these extra Higgs fields also help to satisfy the precision electroweak tests, and other observables. Because the scale of the symmetry breaking is electroweak, both the mirror quark and mirror leptons have masses in the electroweak scale in the range $ \sim 150 - 800 $ GeV. The mirror quarks / leptons  decay to ordinary quarks /leptons plus very light neutral scalars. In this work,  we calculate the final state signals  arising from the pair productions of these mirror quarks and their subsequent decays.  We find that these signals are well observable over the Standard Model background for $13$ TeV LHC. Depending on the  associated Yukawa couplings, these decays can also give rise displaced vertices with long decay length, very different from the usual displaced vertices associated with b decays.


\end{abstract}


\maketitle

\section{Introduction}

Two of the outstanding experimental problems of the highly successful Standard Model (SM) is the existence of the non-zero neutrino masses and the dark matter. One can obtain non-zero neutrino masses in the SM from the effective dimension 5 operators \cite{Weinberg} (Weinberg operator, having the schematic structure of LLHH, where L is a lepton doublet and H the Higgs doublet), but if the Planck scale is the next scale in the theory beyond the $\tev$ scale, then the neutrino masses comes out to few orders of magnitude smaller compared to the observed values. The scale needed is $\sim 10^{14}$ GeV. If there are right handed (RH) SM singlet neutrinos at this scale, then such effective operators can be obtained by integrating out the heavy RH neutrinos.  Seesaw mechanism constructed by postulating such SM singlet RH neutrinos is the most elegant mechanism to explain these tiny neutrino masses. However, existence of such a heavy RH singlet neutrino can not be tested in any laboratory experiments , specially in the currently running high energy large hadron colliders (LHC). Also, fermion representation in the SM is very asymmetric, left handed (LH) fermions are doublets, whereas RH fermions are singlets. Long ago, it was proposed that may be nature is more symmetric,
there are similar heavy particle with exactly opposite chirality \cite{leeyang}. However, such a simple extension  is excluded by the currently available precision electroweak data, namely the S parameter.
Recently, a new mirror symmetric model has been proposed \cite{pqnur} which rectifies the old Lee-Yang proposal by extending the Higgs sector. The  electroweak gauge symmetry is $ SU(2)_W \times U(1)_Y$, and for every left-handed SM doublets , there are right handed SM doublets with new fermions. Similarly for every RH SM singlet, we have new LH singlet fermions. These new fermions are called mirror fermions. The EW precision constraints, such as the S parameter is satisfied by extending the Higgs sector to include $SU(2)$ triplet Higgs. The $SU(2)_W \times U(1)_Y$ symmetry is broken in the same electroweak scale , $\Lambda_{EW} \sim$ 246 GeV.   All the particles get  masses from the spontaneous symmetry breaking in this scale, will have masses 
less than a TeV. Notice also the marked difference with the Left-Right symmetric model \cite{LR} which is characterized by an extra $SU(2)_R$ with right-handed fermions transforming as doublets under that new gauge group. Furthermore, the L-R model contains {\em two} symmetry breaking scales $\Lambda_L$ and $\Lambda_R$ with the latter being completely arbitrary and constrained experimentally from below by the latest LHC data \cite{WR} $M_R \gtrsim 3 \tev$. In the L-R model, the masses of the right-handed neutrinos are proportional to the $SU(2)_R$  scale. 

In this model,  the  RH neutrinos which are  doublet under the $SU(2)$ gauge symmetry will have masses in the electroweak scale. An explicit model was constructed in which seesaw mechanism is realized to obtain tiny neutrino masses with the RH neutrinos at the EW scale \cite{pqnur}. The implications for the model
for neutrino masses and mixings,  precision EW tests, lepton violating rare processes were discussed \cite{hung2,hung3,pqtrinh}. This model,  uses an  $A_4$  \cite{pqtrinh} discrete symmetry.   In addition to the usual Higgs doublet, it has a second Higgs doublet, two Higgs triplets and several Higgs singlets. After the symmetry breaking,  the neutral Higgses mix. One of the neutral Higgs is the recently observed $125$ GeV Higgs boson. The model also has  doubly charged Higgs, singly charged Higgs, as well as additional massive neutral Higgs. A previous analysis of some aspects of this scalar sector can be found in \cite{pqaranda}.

In this work, we explore the collider implications of this model. The model has mirror fermions, RH doublets and LH singlets, particularly the mirror quarks. These particle have masses below $\sim 800$ GeV to satisfy unitarity, and  can be copiously produced at the LHC vis their strong color  interaction. These can only decay to ordinary quarks and essentially massless scalars. Depending on the relevant Yukawa couplings, these mirror quarks may decay immediately, or may have long life. If they decay immediately, we will get final state with high $p_T$ jets, and large missing energy due the escaping light scalars. If they have long life, then they can give rise to  {2-jets which are coming from} displaced vertices. 
{Therefore, the final state signature, in this case, is characterized by 2-jets plus missing transverse momentum ($p_T\!\!\!\!\!\!/~$) where the jets are not pointing towards the primary vertex. For the LHC multi-jets + $p_T\!\!\!\!\!\!/~$ analysis, such jets are usually considered as "Fake jets" not originating from the hard scattering and thus, rejected as beam induced background and cosmic rays \cite{fake1,fake2}.}

Our presentation below are as follows. In section II through V,  we review the model and the formalism, as well as the 
the constraints from the electroweak precision tests  and the Higgs data. Section VI contains our new results
where we discuss the implications of the models at the LHC. We conclude in section VII. 

We end this introduction by mentioning an aspect of the SM which often goes unnoticed. The electroweak phase transition is intrinsically non-perturbative and a common framework for studying non-perturbative phenomena is that of lattice gauge theory. It has been problematic to put a chiral gauge theory such as the SM on the lattice because of the loss of gauge invariance. Ref.~\cite{montvay} proposed the introduction of mirror fermions in order to achieve a gauge-invariant formulation of the SM on the lattice. The mirror fermions of the EW-scale $\nu_R$ model would play such a role.

Finally, we take the liberty to quote a sentence from the famous paper about parity violation of Lee and Yang \cite{leeyang}: "If such asymmetry is indeed found, the question could still be raised whether there could not exist corresponding elementary particles exhibiting opposite asymmetry such that in the broader sense there will still be over-all right-left symmetry.." The EW-scale $\nu_R$ model \cite{pqnur} is a direct response to this famous quote and is a model which satisfies the electroweak precision data as we have mentioned above.

\section{ The model, formalism and the existing constraints}

The EW-scale $\nu_R$ model \cite{pqnur} is basically the SM with an extended fermionic [and scalar] sector: For every SM left-handed doublet, there is a mirror right-handed doublet, and for every SM right-handed singlet, there is a mirror left-handed singlet. The gauge group remains the same and, as such, the energy scale characterizing the EW-scale $\nu_R$ model is still the electroweak scale $\Lambda_{EW} \sim 246 \, \gev$. This is the reason, as we shall see in the brief review, for the Majorana mass of the right-handed neutrinos to be bounded from above by the electroweak scale and, as a consequence, for its accessibility at colliders such as the LHC and the International Linear Collider (ILC). However, it is important to note that as discussed in \cite{hung3} in detail, the statement that the right-handed neutrino masses are bounded by the electroweak scale is assuming $g_M \sim O(1)$.

\begin{itemize}

\item {\bf Gauge group of the EW-scale $\nu_R$ model}:

\be
SU(3)_C \times SU(2)_W \times U(1)_Y
\ee

Notice the absence of the subscript "L" in $SU(2)$. This is because the EW-scale $\nu_R$ model accommodate both SM fermions and the mirror counterparts of opposite chirality. Notice also the marked difference with the Left-Right symmetric model \cite{LR} which is characterized by an extra $SU(2)_R$ with right-handed fermions transforming as doublets under that new gauge group. Furthermore, the L-R model contains {\em two} symmetry breaking scales $\Lambda_L$ and $\Lambda_R$ with the latter being completely arbitrary and constrained experimentally from below by the latest LHC data \cite{WR} $M_R \gtrsim 3 \tev$. In the L-R model, the masses of the right-handed neutrinos are proportional to the $SU(2)_R$ scale.

\item {\bf Fermion $SU(2)_W$ doublets} ($M$ refers to mirror fermions):

SM: $l_L = \left(
	  \begin{array}{c}
	   \nu_L \\
	   e_L \\
	  \end{array}
	 \right)$; Mirror: $l_R^M = \left(
	  \begin{array}{c}
	   \nu_R^M \\
	   e_R^M \\
	  \end{array}
	 \right)$. \
	 
SM: $q_L = \left(
	  	 \begin{array}{c}
	   	  u_L \\
	     	  d_L \\
	  	\end{array}
	 	\right)$; Mirror: $q_R^M = \left(
	  	 \begin{array}{c}
	   	  u_R^M \\
	     	  d_R^M \\
	  	\end{array}
	 	\right)$.

\item {\bf Fermion $SU(2)_W$ singlets}:

SM: $e_R; \ u_R, \ d_R$; Mirror: $e_L^M; \ u_L^M, \ d_L^M$

\item{Scalar sector}
\begin{itemize}
 \item A singlet scalar Higgs $\phi_S$ with $\langle \phi_S \rangle = v_S$.
 In \cite{pqnur}, it was stated that the Dirac mass appearing in the seesaw formula, namely $m_\nu^D = g_{Sl} \ v_S $ (see the review below), has to be less than 100 keV in order for $m_{\nu} < O(eV)$ because $M \sim O(\Lambda_{EW})$. Furthermore, if one assumes $g_{Sl} \sim O(1)$ then $v_S \sim O(100\, keV)$ and the Higgs singlet particle mass will be comparable to that value or smaller and will be much lighter than the other particles. However, an updated analysis of $\mu \rightarrow e \gamma$ \cite{hung4} constrains $g_{Sl} < 10^{-3}$ which gives $v_S \sim O(100\, \mev)$. Nevertheless, one can easily obtain the mass of the physical Higgs singlet scalar to be smaller than that value. In what follows, we could safely ignore the singlet mass in our phenomenological analysis. Notice that the aforementioned statements are independent of the values of the Yukawa couplings $g_{Sq}$ present in the SM-Mirror-singlet Higgs interactions which are not constrained by experiment at this moment.
 
	 \item Doublet Higgses:
	 
	  $\Phi_2=\left(
	 			\begin{array}{c}
				\phi_{2}^+ \\
				\phi_{2} ^0 \\
				\end{array}
				\right)$ 		
with $\langle \phi_{2}^0 \rangle = v_2/\sqrt{2}$. 

In the original version \cite{pqnur}, this Higgs doublet couples to both SM and mirror fermions. An extended version was proposed \cite{hung3} in order to accommodate the 125-GeV SM-like scalar and, in this version, $\Phi_2$ only couples to SM fermions while another doublet $\Phi_{2M}$ whose VEV is $\langle \phi_{2M}^0 \rangle = v_{2M}/\sqrt{2}$ couples only to mirror fermions. 

	 \item Higgs triplets
	 	\begin{itemize}
			\item $\widetilde{\chi} \ (Y/2 = 1)  = \frac{1}{\sqrt{2}} \ \vec{\tau} . \vec{\chi} = 
	  \left(
	  \begin{array}{cc}
	    \frac{1}{\sqrt{2}} \chi^+ & \chi^{++} \\
	    \chi^0 & - \frac{1}{\sqrt{2}} \chi^+\\
	   \end{array}
		  \right)$ with $\langle \chi^0 \rangle = v_M$.
			\item $\xi \ (Y/2 = 0)$ in order to restore Custodial Symmetry with $\langle \xi^0 \rangle = v_M$.
			
	\item VEVs:
	
	$v_{2}^2 +v_{2M}^2 + 8 v_{M}^2= v^2 \approx (246 \gev)^2$  		
		\end{itemize}	
\item {\bf Dirac and Majorana Neutrino Masses}
For simplicity, from hereon, we will write $\nu_R^M$ simply as $\nu_R$.
\begin{itemize}
\item Dirac Neutrino Mass\\
The singlet scalar field $\phi_S$ couples to fermion bilinear
\bea
\label{dirac}
L_S &=& g_{Sl} \,\bar{l}_L \ \phi_S \ l_R^M + h.c.\\  \nonumber
         &= &g_{Sl} (\bar{\nu}_L \ \nu_R \ + \bar{e}_L \ e_R^M) \ \phi_S + h.c. \,.
\eea
From (\ref{dirac}),  we get the Dirac neutrino masses $m_\nu^D = g_{Sl} \ v_S $.\\
\item Majorana Neutrino Mass \\
\bea
\label{majorana}
L_M &= &g_M \, l^{M,T}_R \ \sigma_2 \ \tau_2 \ \tilde{\chi} \ l^M_R \\ \nonumber
&= &g_M \ \nu_R^T \ \sigma_2 \ \nu_R \ \chi^0 - \dfrac{1}{\sqrt{2}} \ \nu_R^T \ \sigma_2 \ e_R^M \ \chi^+ \\ \nonumber
&&- \dfrac{1}{\sqrt{2}} \ e_R^{M,T} \ \sigma_2 \ \nu_R \ \chi^+ + e_R^{M,T} \ \sigma_2 \ e_R^M \ \chi^{++} \,.
\eea
From (\ref{majorana}), we obtain the Majorana mass $ M_R = g_M v_M $.
\end{itemize}
\end{itemize}
\end{itemize}
It is important to note here that in the original version \cite{pqnur}, a global symmetry denoted by $U(1)_M$ was assumed under which the mirror right-handed doublets and left-handed singlets transform as $(l_R^{M}, e_L^{M}) \rightarrow e^{\imath \theta_M} (l_R^{M}, e_L^{M})$ and the triplet and singlet Higgs fields transform as $\tilde{\chi} \rightarrow e^{-2\imath \theta_M} \tilde{\chi}$, $\phi_S \rightarrow e^{-\imath \theta_M} \phi_S$, with all other fields being singlets under $U(1)_M$. With this transformation, a coupling similar to Eq  (\ref{majorana}) is forbidden for the SM leptons and hence there is no Majorana mass for left-handed neutrinos at tree level. It was also shown in \cite{pqnur} that the Majorana mass for left-handed neutrinos can arise at one loop but is much smaller than the light neutrino mass and thus can be ignored.

The next section will be devoted to a review of results which have been obtained from the EW-scale $\nu_R$ model \cite{hung2,hung3}. Since the previous section and the one that follows are necessary to introduce the model to readers who are not familiar with the model and, in particular, its phenomenological consequences, we include similar reviews in all related papers.

\section  {Electroweak precision constraints on the EW $\nu_R$ model \cite{hung2}}

The presence of mirror quark and lepton $SU(2)$-doublets can, by themselves, seriously affect the constraints coming from electroweak precision data. As noticed in \cite{pqnur}, the positive contribution to the S-parameter coming from the extra right-handed mirror quark and lepton doublets could be partially cancelled by the negative contribution coming from the triplet Higgs fields. Ref.~\cite{hung2} has carried out a detailed analysis of the electroweak precision parameters S and T and found that there is a large parameter space in the model which satisfies the present constraints and that there is {\em no fine tuning} due to the large size of the allowed parameter space. It is beyond the scope of the paper to show more details here but a representative plot would be helpful. Fig. 1 shows the contribution of the scalar sector versus that of the mirror fermions to the S-parameter within 1$\sigma$ and 2$\sigma$.
\begin{figure}[t]
\centering
    \includegraphics[scale=0.35]{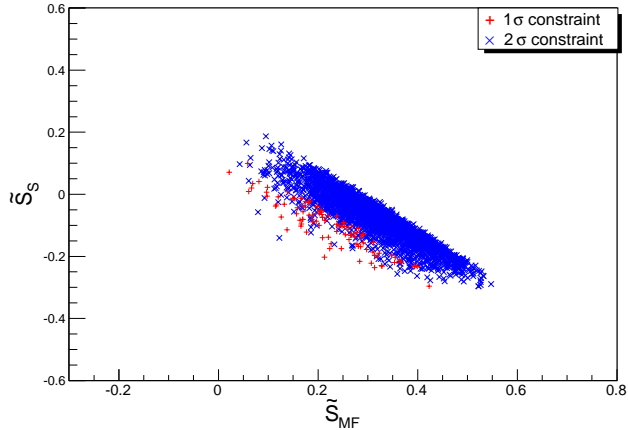} 
 \caption{{\small The plot shows the contribution to the S-parameter for the scalar sector ($\tilde{S}_S$) vs the mirror fermion sector ($\tilde{S}_{MF}$) within the 1 and 2 $\sigma$'s allowed region. The negative contribution to the S-parameter from the scalar sector tends to partially cancel the positive contribution from the mirror fermion sector and the total sum of the two contributions agrees with experimental constraints. }}
\label{SsvsSmf}
\end{figure}
In the above plot, \cite{hung2} took for illustrative purpose 3500 data points that fall inside the 2$\sigma$ region with about 100 points falling inside the 1$\sigma$ region. More details can be found in \cite{hung2}. 

\section {Review of the scalar sector of the EW $\nu_R$ model in light of the discovery of the 125-GeV SM-like scalar \cite{hung3}}

In light of the discovery of the 125-GeV SM-like scalar, it is imperative that any model beyond the SM (BSM) shows a scalar spectrum that contains at least one Higgs field with the desired properties as required by experiment. The present data from CMS and ATLAS only show signal strengths that are compatible with the SM Higgs boson. The definition of a signal strength $\mu$ is as follows 
\be
\sigma(H \text{-decay}) = \sigma(H \text{-production}) \times BR(H \text{-decay})\,,
\ee
and
\be\label{eq:mudef}
	\mu(H \text{-decay}) = \frac{\sigma(H \text{-decay})}{\sigma_{SM}(H \text{-decay})}\,.
\ee

To really distinguish the SM Higgs field from its impostor, it is necessary to measure the partial decay widths and the various branching ratios. In the present absence of such quantities, the best one can do is to present cases which are consistent with the experimental signal strengths. This is what was carried out in \cite{hung3}. 

The minimization of the potential containing the scalars shown above breaks its global symmetry $SU(2)_L \times SU(2)_R$ down to a custodial symmetry $SU(2)_D$ which guarantees at tree level $\rho = M_{W}^2/M_{Z}^2 \cos^2 \theta_W=1$ \cite{hung3}. The physical scalars can be grouped, based on their transformation properties under $SU(2)_D$ as follows:
	\begin{eqnarray}
		\text{five-plet (quintet)} &\rightarrow& H_5^{\pm\pm},\; H_5^\pm,\; H_5^0;\nonumber\\[0.5em]
		\text{triplet} &\rightarrow& H_{3}^\pm,\; H_{3}^0;\nonumber\\[0.5em]
		\text{triplet} &\rightarrow& H_{3M}^\pm,\; H_{3M}^0;\nonumber\\[0.5em]
		\text{three singlets} &\rightarrow& H_1^0,\; H_{1M}^0,\; H_1^{0\prime}\,,
	\end{eqnarray}
  The three custodial singlets are the CP-even states, one combination of which can be the 125-GeV scalar. In terms of the original fields, one has $H_1^0 = \phi_{2}^{0r}$,  $H_{1M}^0 = \phi_{2M}^{0r}$, and $H_1^{0\prime} = \frac{1}{\sqrt{3}} \Big(\sqrt{2}\chi^{0r}+ \xi^0\Big)$. These states mix through a mass matrix obtained from the potential and the mass eigenstates are denoted by $\widetilde{H}$, $\widetilde{H}^\prime$, and $\widetilde{H}^{\prime\prime}$, with the convention that the lightest of the three is denoted by $\widetilde{H}$, the next heavier one by $\widetilde{H}^\prime$ and the heaviest state by $\widetilde{H}^{\prime\prime}$. 
  
  To compute the signal strengths $\mu$, Ref.~\cite{hung3} considers $\widetilde{H} \rightarrow ZZ,~W^+W^-,~\gamma\gamma,~b\bar{b},~\tau\bar{\tau}$. In addition, the cross section of $g g \rightarrow \widetilde{H}$ related to $\widetilde{H} \rightarrow g g$ was also calculated. A scan over the parameter space of the model yielded {\em two interesting scenarios} for the 125-GeV scalar: 1) {\em Dr Jekyll}'s scenario in which $\widetilde{H} \sim H_1^0$ meaning that the SM-like component $H_1^0 = \phi_{2}^{0r}$ is {\em dominant}; 2) 
{\em Mr Hyde}'s scenario in which $\widetilde{H} \sim H_1^{0\prime}$ meaning that the SM-like component $H_1^0 = \phi_{2}^{0r}$ is {\em subdominant}. Both scenarios give signal strengths compatible with experimental data as shown below in Fig.~(2).
\begin{figure}
\label{signal2}
	\centering
	\includegraphics[width=0.5\textwidth]{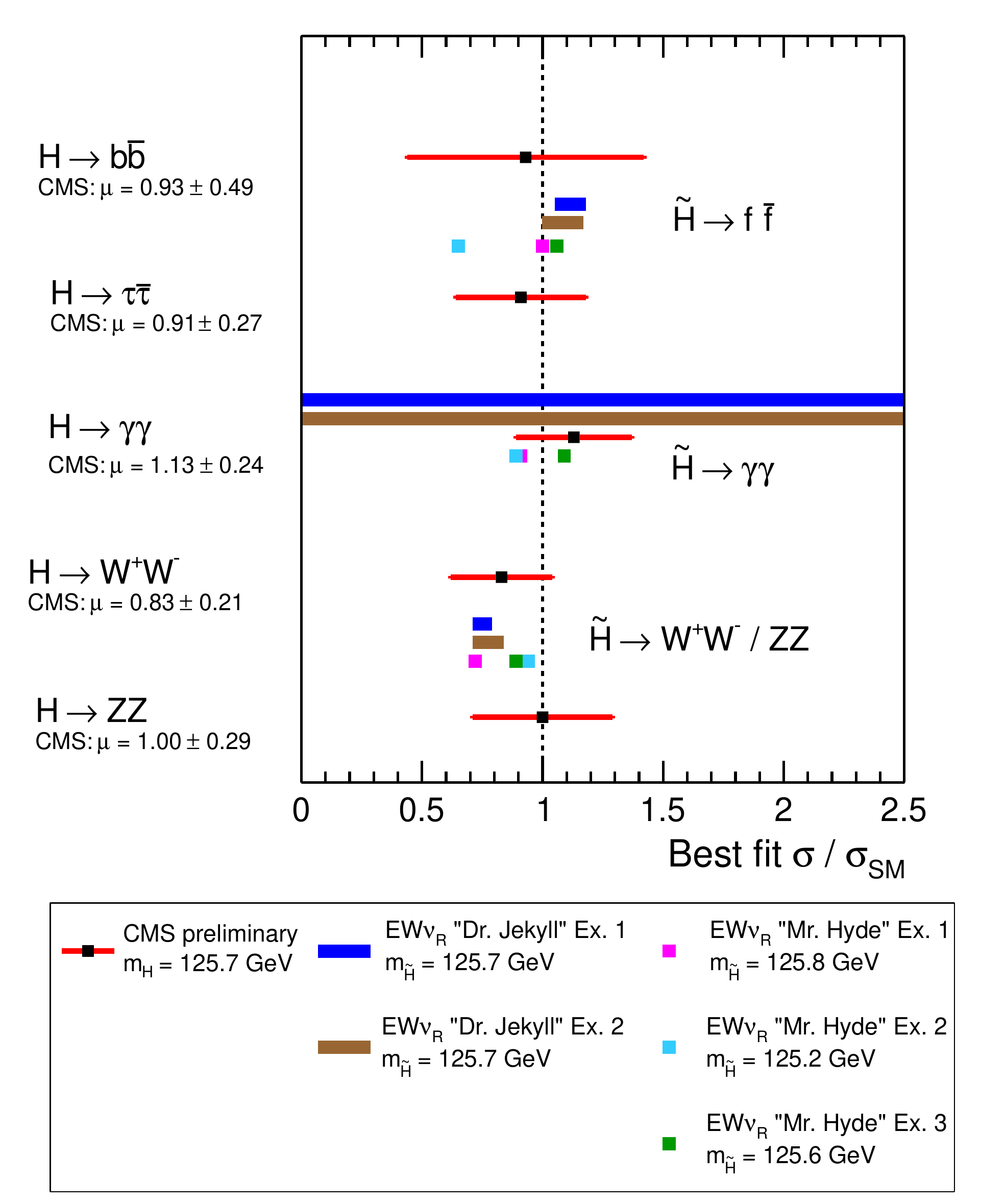}
	\caption{Figure shows the predictions of $\mu(\widetilde{H} \rightarrow ~b\bar{b}, ~\tau\bar{\tau}, ~\gamma\gamma, ~W^+W^-, ~ZZ)$ in the EW $\nu_R$ model for examples 1 and 2 in {\em Dr.~Jekyll} and example 1, 2 and 3 in {\em Mr.~Hyde} scenarios as discussed in \cite{hung3}, in comparison with corresponding best fit values by CMS \cite{h_ww_122013, h_zz_4l_122013, h_bb_102013, h_tautau_012014}.}
\end{figure}

As we can see from Fig.~(2), both SM-like scenario ({\em Dr Jekyll}) and the {\em more interesting scenario} which is very unlike the SM ({\em Mr Hyde}) agree with experiment. As stressed in \cite{hung3}, present data cannot tell whether or not the 125-GeV scalar is truly SM-like or even if it has a dominant SM-like component. It has also been stressed in \cite{hung3} that it is essential to measure the partial decay widths of the 125-GeV scalar to truly reveal its nature. Last but not least, in both scenarios, $H_{1M}^0 = \phi_{2M}^{0r}$ is subdominant but is essential to obtain the agreement with the data as shown in \cite{hung3}.

As discussed in detail in \cite{hung3} , for proper vacuum alignment, the potential contains a term proportional to $\lambda_5$ (Eq.~(32) of \cite{hung3}) and it is this term that prevents the appearance of Nambu-Goldstone (NG) bosons in the model. The would-be NG bosons acquire a mass proportional to $\lambda_5$ .

An analysis of CP-odd scalar states $H_{3}^0, H_{3M}^0 $ and the heavy CP-even states $\widetilde{H}^\prime$, and $\widetilde{H}^{\prime\prime}$ was presented in \cite{hung3}. The phenomenology of charged scalars including the doubly-charged ones was also discussed in \cite{pqaranda}.

The phenomenology of mirror quarks and leptons was briefly discussed in \cite{pqnur} and a detailed analysis of mirror quarks will be presented in this paper. It suffices to mention here that mirror fermions decay into SM fermions through the process $q^M\rightarrow q\phi_S$, $l^M\rightarrow l\phi_S$ with $\phi_S$ "appearing" as missing energy in the detector. Furthermore, the decay of mirror fermions into SM ones can happen outside the beam pipe and inside the silicon vertex detector. Searches for non-SM fermions do not apply in this case. It is beyond the scope of the paper to discuss these details here.

\section{Yukawa interactions between mirror and SM quarks}

\begin{itemize}

\item {\bf The interactions}:

The EW $\nu_R$ model has been extended to include an investigation of neutrino and charged lepton mass matrices and mixings \cite{pqtrinh}. In \cite{pqtrinh}, a non-abelian discrete symmetry group $A_4$ was assumed and was applied to the Higgs singlet sector which is responsible for the Dirac masses of the neutrinos. Following \cite{pqtrinh}, we list the assignments of the SM and mirror fermions as well as those for the scalars under $A_4$.
\begin{table}[!htb]
\caption{\label{assignment} $A_4$ assignments for leptons and Higgs fields}
\begin{center}
\begin{tabular}{| c || c | c | c | c | c | c | c |}
 \hline
 Field & $\mathnormal{(\nu,l)_L}$ & $\mathnormal{(\nu, l^M)_R}$ & $\mathnormal{e_R}$ & $\mathnormal{e_L^M}$ & $\mathnormal{\phi_{0S}}$ & $\mathnormal{\tilde{\phi}_{S}}$ & $\mathnormal{\Phi_2}$ \\ [0.5ex]
 \hline
 $A_4$ & $\underline{3}$ & \underline{3} & \underline{3} & \underline{3} & \underline{1} & \underline{3} & \underline{1}\\
  \hline
\end{tabular}
\end{center}
\end{table} 
From this assignment, one obtains the following Yukawa interactions in terms of quark mass eigenstates ($q^{d}_L=(d_L,s_L,b_L)$, $q^{u}_L=(u_L,c_L,t_L)$, $q^{M,d}_{R}=(d^{M}_R, s^{M}_R, b^{M}_R)$, $q^{M,u}_{R}=(u^{M}_R, c^{M}_R, t^{M}_R)$):
\bea
\label{yukawadown}
L_S&=& \bar{q}^{d}_{L}\, U^{d\dagger}_{L}  M^{d}_{\phi}  U^{d^M}_R \, q^{M,d}_{R} +H.c. \; \; \nonumber \\
        &=& \bar{q}^{d}_{L}\, \bar{M^{d}_{\phi}}  \, q^{M,d}_{R} +H.c. \; 
\eea
for the down quark sector and
\bea
\label{yukawaup}
L_S&=& \bar{q}^{u}_{L}\, U^{u\dagger}_{L}  M^{u}_{\phi}  U^{u^M}_R \, q^{M,u}_{R} +H.c. \; \; \nonumber \\
        &=& \bar{q}^{u}_{L}\, \bar{M^{u}_{\phi}}  \, q^{M,u}_{R} +H.c. \; 
\eea
for the up quark sector and where
\be
\label{mnu}
M^{d,u}_{\phi} = 
\left(
  \begin{array}{cccc}
    g^{d,u}_{0S}\phi_{0S} & g^{d,u}_{1S}\phi_{3S} & g^{d,u}_{2S}\phi_{2S} \\
    g^{d,u}_{2S}\phi_{3S} & g^{d,u}_{0S}\phi_{0S}  & g^{d,u}_{1S}\phi_{1S} \\
    g^{d,u}_{1S}\phi_{2S}  & g^{d,u}_{2S}\phi_{1S} & g^{d,u}_{0S}\phi_{0S}  \\
  \end{array}
\right) \, .
\ee

The mixing parameters involving in the decay $q^{M,i}_R \rightarrow q^{j}_L + \phi_l$ where $i$ and $j$ denote quark flavors and $l=0,..,3$ are contained in the parametrizations of $\bar{M^{d,u}_{\phi}}$ as well as in Eq.~(\ref{mnu}).

An important remark is in order here. Unlike the Yukawa couplings $g_{Sl}$ of the lepton sector which are constrained by rare processes such as $\mu \rightarrow e \, \gamma$, no such constraint exists for $g^{d,u}_{iS}$ and they can be arbitrarily small as the present upper bounds on BR($t \rightarrow q Z$) from CMS and ATLAS are $5 \times 10^{-4}$ and $7 \times 10^{-4}$ respectively and are not ``low" enough to constrain the Yukawa couplings $g_{Sq}$.  This can give rise to displaced vertices of the type shown in Fig.~3. For this kind of rare decay modes such as $t \rightarrow q Z$, etc it should exist through a loop diagram as the ones calculated in \cite{hung4} for $\mu \rightarrow e \gamma$. These processes are indeed under investigation by the authors of \cite{hung4} and the results indicate values for BR which are many orders of magnitudes smaller than the current limits even when $g_{Sq} \sim O(1)$. 

\item {\bf The decay width}:
 
Since we will be concentrating below on the production and signature of the lightest mirror quark, the decay mode that is allowed is $q^M\to q \phi_S~{\rm or}~b \phi_S$. As stressed in \cite{pqnur}, the singlet scalars are assumed to be much lighter than the quarks (both SM and mirror) and we will neglect their masses in the computation of the decay width. One obtains
\bea
\label{decaywidth}
\Gamma(q^M\rightarrow q+\phi^{\star})~=~\dfrac{g_{Sq}^2}{64\pi}m_{q^M}\left(1-\frac{m_q^2}{m_{q^M}^{2}}\right)\left(1+\frac{m_q}{m_{q^M}}-\frac{m_q^2}{2m_{q^M}^2}\right) \, ,
\eea
where explicit expressions for the generic coupling $g_{Sq}$ in Eq.~(\ref{decaywidth}) can be obtained by using Eqs.~(\ref{yukawadown}, \ref{yukawaup}, \ref{mnu}). In $g_{Sq}$, one finds the Yukawa coupling and various mixing angles. Since the decay length is $\gamma \, \beta \, \hbar \,c/\Gamma(q^M\rightarrow q+\phi^{\star})$, one easily imagine that it can be {\em macroscopic} i.e. $> 1\, mm$ if $g_{Sq}$ is sufficiently small. In Fig.~3 the variation of the decay length (cm) with the generic coupling $g_{Sq}$ varying in between $10^{-8}$ to $10^{-7}$ is shown. For these values of the coupling $g_{Sq}$, the decays of these lightest mirror quarks can produce displaced vertices with decay length varying in the range of few mm to few cm, which can be easily distinguished from the displaced vertices produced by b quarks having average decay length of $\sim 0.5$ mm.
\begin{figure}
\label{decaylength}
	\centering
	\includegraphics[width=0.5\textwidth]{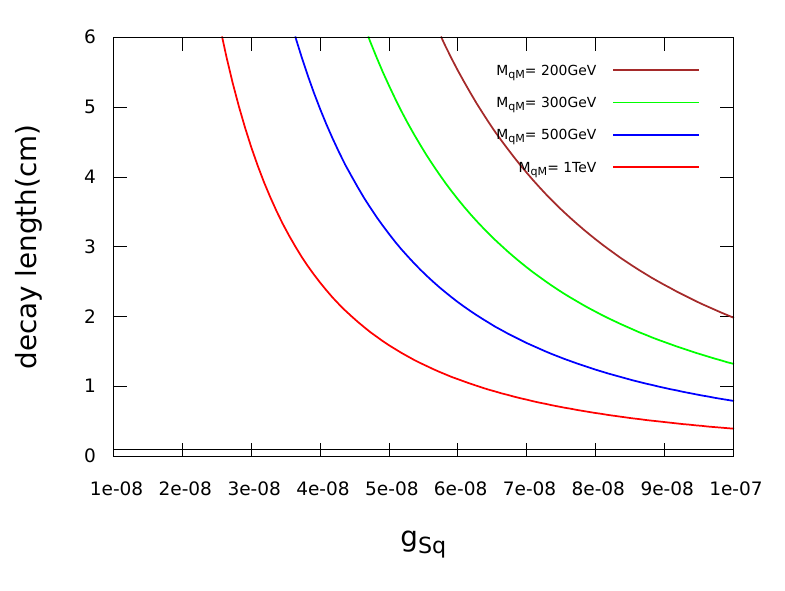}
	\caption{Figure shows the variation of the decay length (cm) with the generic coupling $g_{Sq}$ varying in the range of $10^{-8}$ to $10^{-7}$ for four different values of the mirror quark masses ($M_{q^M} = 200, 300, 500$ and $1000$ GeV). The black horizontal line in the plot corresponds to the decay length of 1mm. So any macroscopic decay length of the mirror quarks above the black line is significantly different than the decay length coming from the b-quarks displaced vertices.}
\end{figure}
Such macroscopic decay length can presently be missed due to the nature of the algorithms of the LHC detectors CMS and ATLAS.
\end{itemize}

\section{Phenomenology: new physics at the LHC}
In this section, we will discuss the collider signature of this model in the context of the LHC experiment. The LHC is a proton proton collider. Therefore, the strongly interacting particles are copiously produced [if kinematically accessible]. In this work, we have studied the production and signature of mirror quarks. Mirror quarks are pair produced at the LHC and pair production takes place via gauge interaction only. At the partonic level, the $gg \to q^M \bar q^M$ process is mediated by a gluon in the $s$-channel or a mirror quark in the $t$-channel, whereas, $q \bar q \to q^M \bar q^M$ process is only mediated by a 
gluon in the $s$-channel. While the electroweak diagrams also contribute to $q \bar q \to q^M \bar q^M$, these contributions are suppressed by a relative factor of 
$(\alpha_{EW}/\alpha_s)^2$. Given that they do not bring in any new topologies, their contributions are  subdominant. Due to larger production cross section, we have studied the production and signature of the lightest mirror quark. Being lightest, it can only decay into a SM quark (light quark or $b$-quark) and the singlet scalar $\phi_S$: $q^M\to q \phi_S~{\rm or}~b \phi_S$. Here, we have assumed that the $d^{M}$ is the lightest mirror quark, making the decay products $b \phi_S$ or $(s,d) \phi_S$.

\begin{figure}
	\centering
	\includegraphics[width=0.45\textwidth, angle=0]{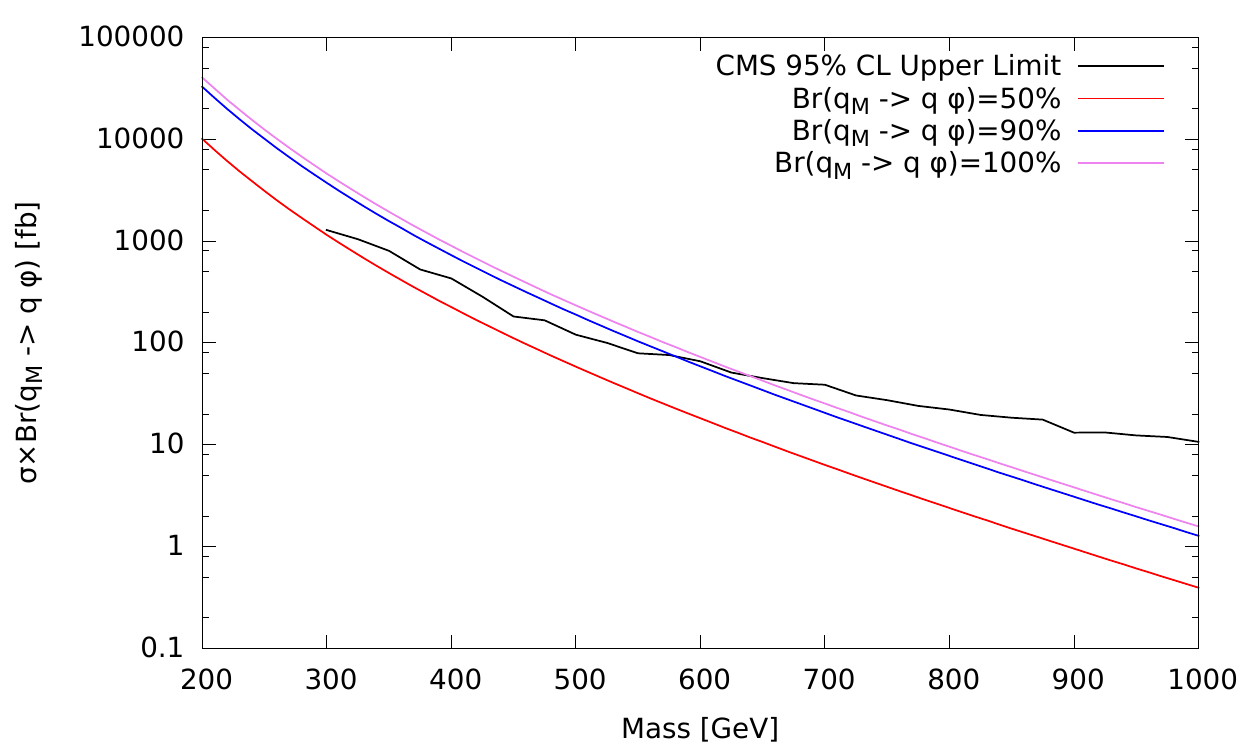}
	\includegraphics[width=0.45\textwidth, angle=0]{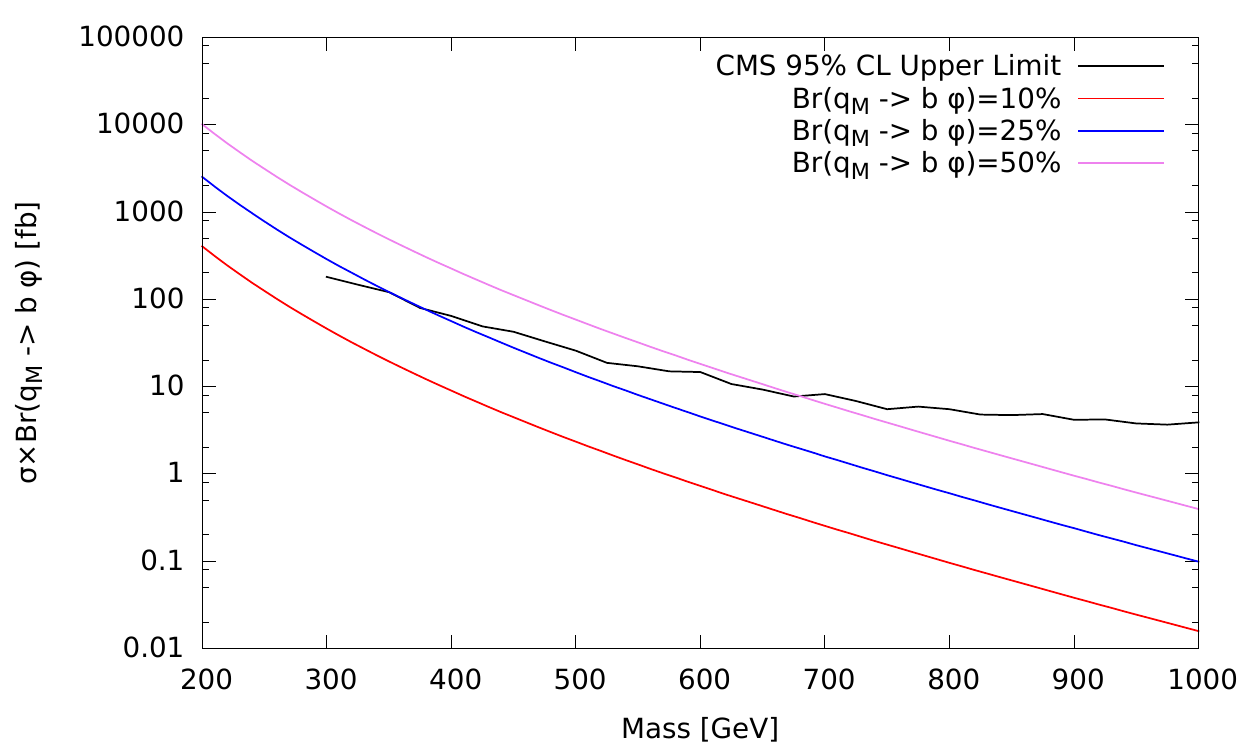}
	\caption{Black line corresponds to 95\% CL upper limits on the theory production cross-section times branching ratio into jets + $p_T\!\!\!\!\!\!/~$ (left panel) and b-jets + $p_T\!\!\!\!\!\!/~$ (right panel) obtained by the CMS group \cite{CMS_jets} with 8 TeV center of mass energy and 11.7 fb$^{-1}$ integrated luminosity. Other lines corresponds to our model prediction for the production cross-section times branching ratio for different values of the branching ratio.}
\label{exclusion}
\end{figure}

These decays take place via Yukawa interactions. Since these Yukawa couplings are free parameter in this model, the decay branching ratios of a mirror quark into a light quark or a $b$-quark are arbitrary. Therefore, the pair production of the lightest mirror quark at the LHC gives rise to two high transverse momentum jets (light quark jet or $b$-jet) in association with large missing transverse energy ($p_T\!\!\!\!\!\!/~$). 
\begin{equation}
pp \to q^M \bar q^M \to q \bar q + p_T\!\!\!\!\!\!/~~{\rm or}~b \bar b + p_T\!\!\!\!\!\!/
\label{signal}
\end{equation}
The missing transverse momentum arises from the very light singlet scalars $\phi_S$  which remain invisible in the detector. Before going into the analysis of signal and background in the context of LHC run II with 13 TeV center of mass energy, it is important to discuss LHC 8 TeV bounds on this model. There is no dedicated study available from ATLAS or CMS collaboration in the context of the present model. However, the main signatures of this model namely, jets + $p_T\!\!\!\!\!\!/~$ or 2-b + $p_T\!\!\!\!\!\!/~$, have already been studied by ATLAS \cite{ATLAS_jets} and CMS \cite{CMS_jets,Khachatryan:2015wza} collaborations in the context of supersymmetry.  The analysis of the CMS collaboration \cite{CMS_jets} is based on the data
collected by the CMS detector in proton-proton collision at $\sqrt s=8$ TeV with an integrated luminosity of 11.7 fb$^{-1}$. The observed  jets + $p_T\!\!\!\!\!\!/~$ or 2-b + $p_T\!\!\!\!\!\!/~$ data is consistent with the SM background
prediction. The absence of any excess of such events was then translated to an upper bound on the production cross-section times branching ratio of any beyond SM process which gives rise to similar signature. In our analysis, we have used the bounds from Ref.~\cite{CMS_jets} to impose constraints on the mirror quark mass and branching ratios to light jets and b-jet. In Fig.~\ref{exclusion}, we have presented (the black lines) 95\% CL upper limits on the theory production cross-section times branching ratio into jets + $p_T\!\!\!\!\!\!/~$ (left panel) and b-jets + $p_T\!\!\!\!\!\!/~$ (right panel) obtained by the CMS group \cite{CMS_jets} with 8 TeV center of mass energy and 11.7 fb$^{-1}$ integrated luminosity. Fig.~\ref{exclusion} also shows our model prediction for the production cross-section times branching ratio for different values of the branching ratio. Fig.~\ref{exclusion} (left panel) shows that for large $q_M \to q \phi_S$ branching ratio, mirror quark mass below about 600 GeV is excluded. Whereas, if the branching ratio of mirror quark into a light quark is below 50\% then there is no bound on the mirror quark mass.   {Similarly, if the branching ratio of $q_M \to b \phi_S$ is small then there is no bound from Ref.~\cite{CMS_jets}. It is important to note that these bounds  are only applicable when mirror quarks decay at the hard scattering point {\em i.e.,} only for large decay widths of the mirror quarks. However, as discussed in the previous section, in the context of this model, the mirror quark decay width could be small enough for the hadronization of the mirror quarks and displaced decay of the hadronized mirror quarks.}  {
Such an event is not reconstructed by the present  {LHC multi-jets search} algorithm and can be missed. In this case, the above lower bounds on mirror quark masses  {are not applicable} 
}

After discussing the LHC 8 TeV bounds on the mirror quark mass, we are now equipped enough to discuss the phenomenology of this model in the context of the LHC  run II with 13 TeV center of mass energy.

 Several SM processes constitute potential backgrounds for the signal of Eq.(\ref{signal}) and we now discuss the dominant ones in succession.
\begin{itemize}
\item An irreducible background arises from the production of a
  $Z$-boson in association with  multiple jets.  The $Z$-boson decays invisibly and gives rise to the missing transverse  energy signature: 
 \be 
    pp ~\to~ Z+n \mbox{--jets} ~\to~ \nu \bar \nu 
        +n \mbox{--jets} 
\ee 
We use the {\sc Alpgen} \cite{Mangano:2002ea}
generator to estimate the $Z$+jets (upto 3-jets) background contribution.  Although the total cross section for this
  process is very large, the imposition of sufficiently strong $p_T$
  and rapidity requirements on the jets serves to
  suppress it strongly. 
It is important to note that our analysis will not be 
  very sophisticated, it is quite conceivable that we might
  underestimate the background, especially where jet reconstruction is
  concerned.
\begin{figure}
	\centering
	\includegraphics[width=0.45\textwidth, angle=0]{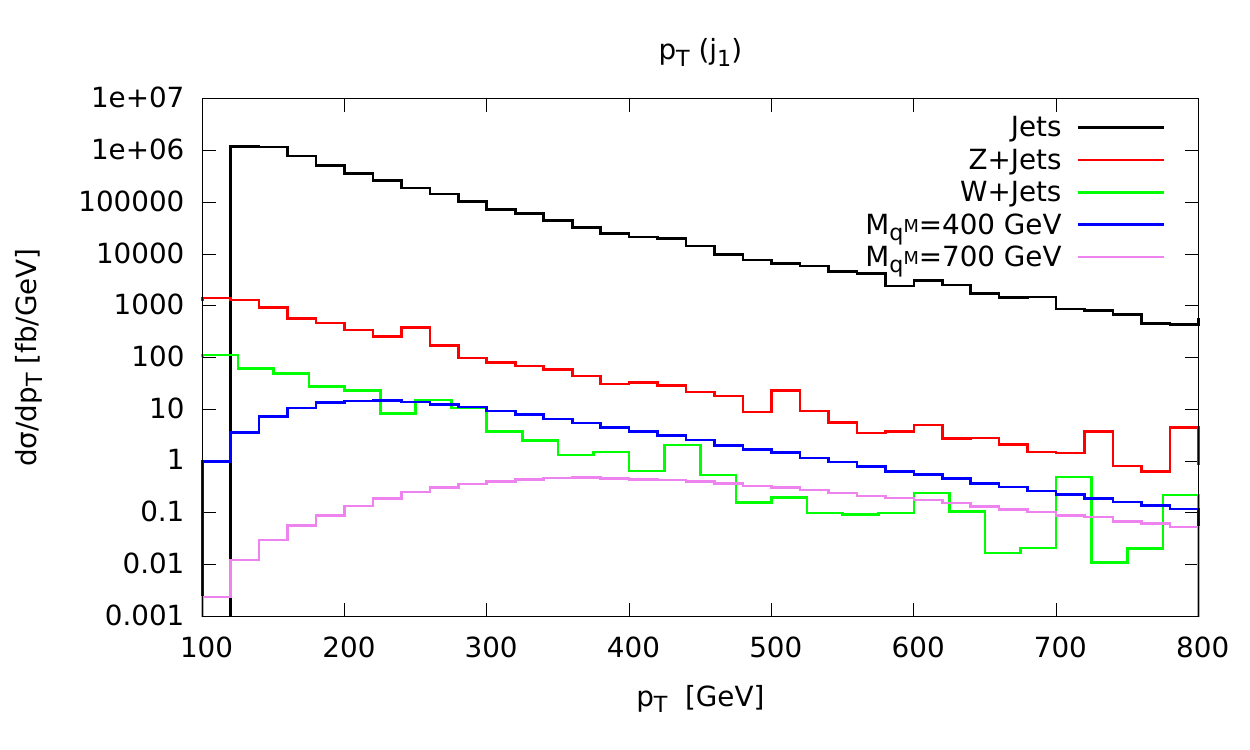}
	\includegraphics[width=0.45\textwidth, angle=0]{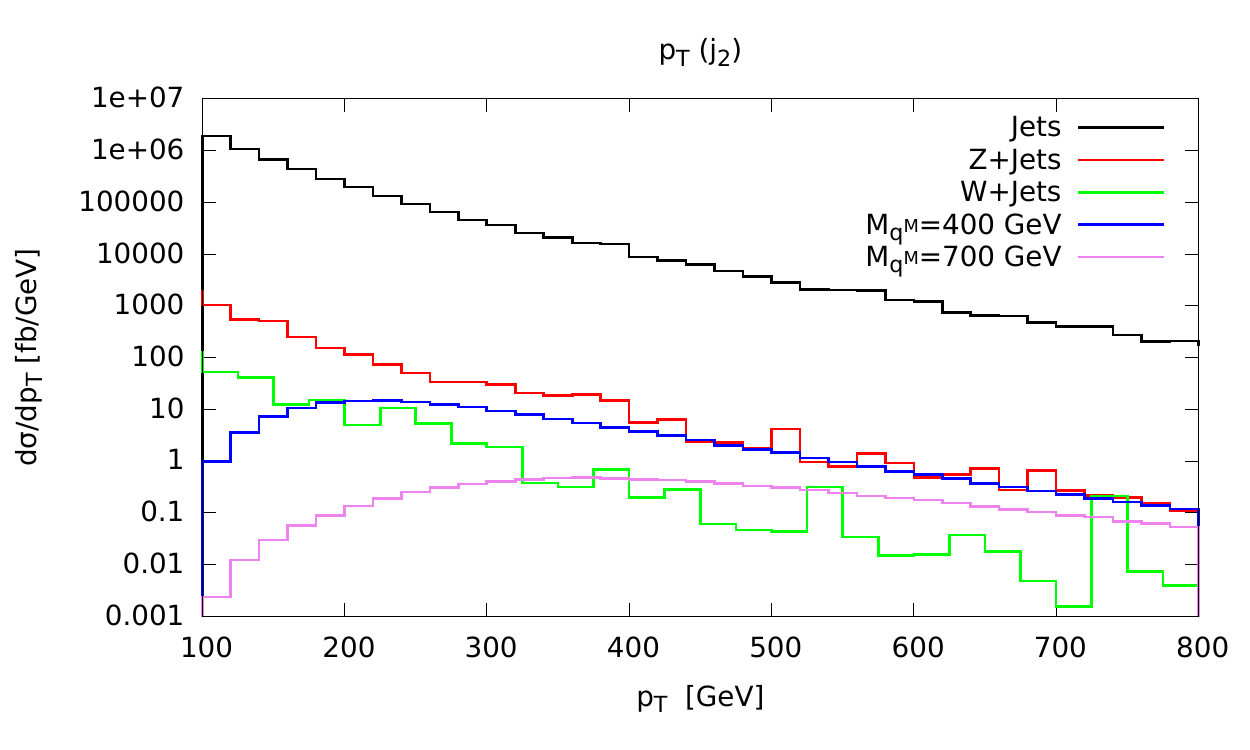}
	\caption{Transverse momentum distributions of jets (left panel: hardest jet; right panel: second hardest jet) after ordering them according to their $p_T$ hardness ($p_T(j_1)>p_T(j_2)$) for signal and background. We have assumed 13 TeV center-of-mass energy of proton-proton collision.}
\label{pt_jet}
\end{figure}

\item The production of $W^\pm$ in association with multiple jets (upto 3-jets) can also be a possible source of background, if the $W^\pm$ decays leptonically and the charged lepton is missed somehow. To be specific, we consider the leptons to be undetectable if it either falls outside the rapidity coverage ($|\eta|\ge 2.5$) or if it is too soft ($p_T\le 20$ GeV) or if it lies too close to any of the jets. In this case, the neutrino and the  missing lepton together give rise to the missing transverse momentum. We also estimate this background too  using {\sc Alpgen}.  Given the fact that the $W$ has a substantial mass and that it is produced with relatively low rapidity, it stands to reason that the charged lepton would, most often, be well within the detector and also have sufficient $p_T$ to be detectable. Consequently, the probability of missing the charged lepton is small, and this background would be suppressed considerably.
\item Significant background contribution can come from the production of multiple jets: $pp~\to~ nj$. In this case, there is no real source of missing transverse momentum. However,  mis-measurement of the $p_T$ of jets can lead to some amount of missing transverse momentum. Since the cross section for the aforementioned process is huge, this process, in principle, could contribute significantly to the background. In this case also, we have used {\sc Alpgen} to compute multi-jets (upto 6-jets) background.
\item The production of $t \bar t$ pairs in association with a $W$ or $Z$-boson ($W$ or $Z$) followed by the leptonic decay of the $W$ or invisible decay of the $Z$ also gives rise to jets + $E_T\!\!\!\!\!\!/~$ background.  In this case, neutrinos in the decay of $W$ or $Z$ gives rise to the missing energy.  The production of $t\bar t+Z/W$ contributes significantly to jets + $E_T\!\!\!\!\!\!/~$ background for higher jet multiplicity since hadronic decay of $t \bar t$ gives rise to 6-jets at parton level. In fact, $t\bar t+Z/W$ is the dominant background for jets + $E_T\!\!\!\!\!\!/~$ signature  for jet multiplicity greater than 4 (see Table~5 of Ref.~\cite{ATLAS_jets}). However, for low jet multiplicity $W/Z$ + jets is the dominant background for jets + $E_T\!\!\!\!\!\!/~$ signature after selection cuts are introduced. For example, if we consider events with at least 2-jets + $E_T\!\!\!\!\!\!/~$ (which is the signal under consideration in this paper), $W/Z$ + jets contribution is at least 10 times bigger than $t\bar t+Z/W$ contribution \cite{ATLAS_jets}.
\end{itemize}

\begin{figure}
	\centering
	\includegraphics[width=0.45\textwidth, angle=0]{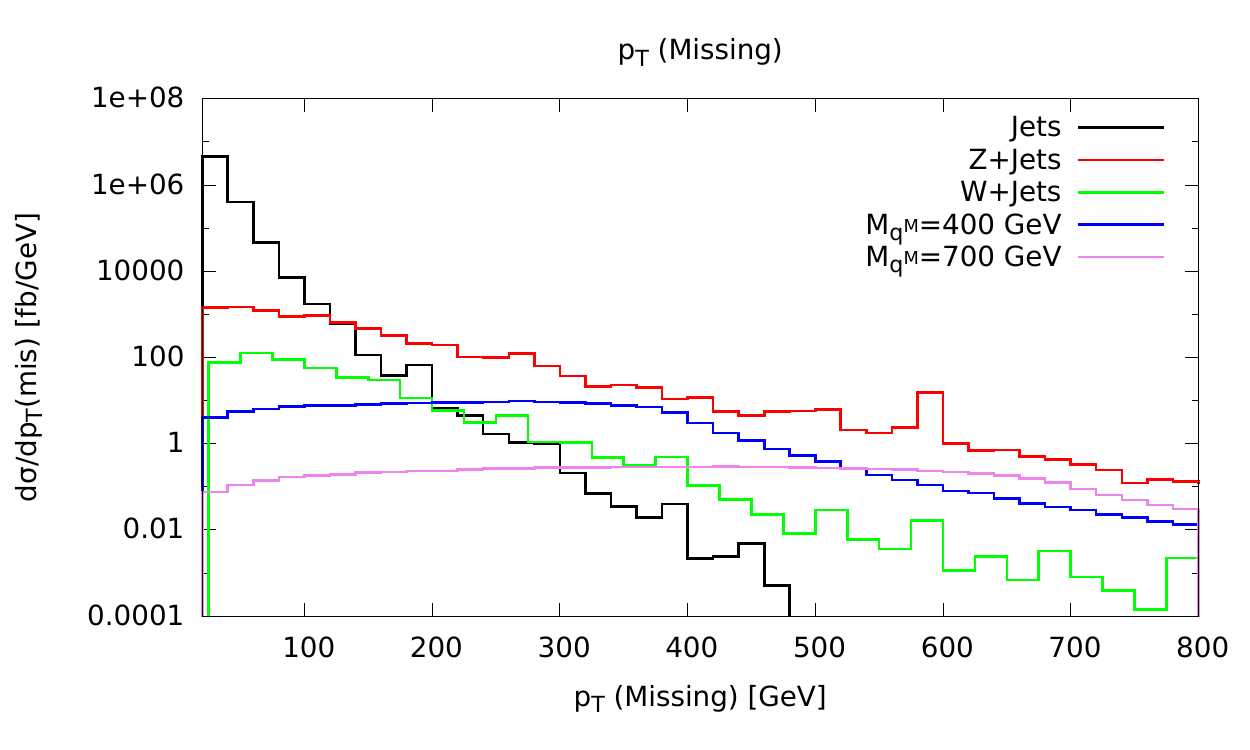}
	\includegraphics[width=0.45\textwidth, angle=0]{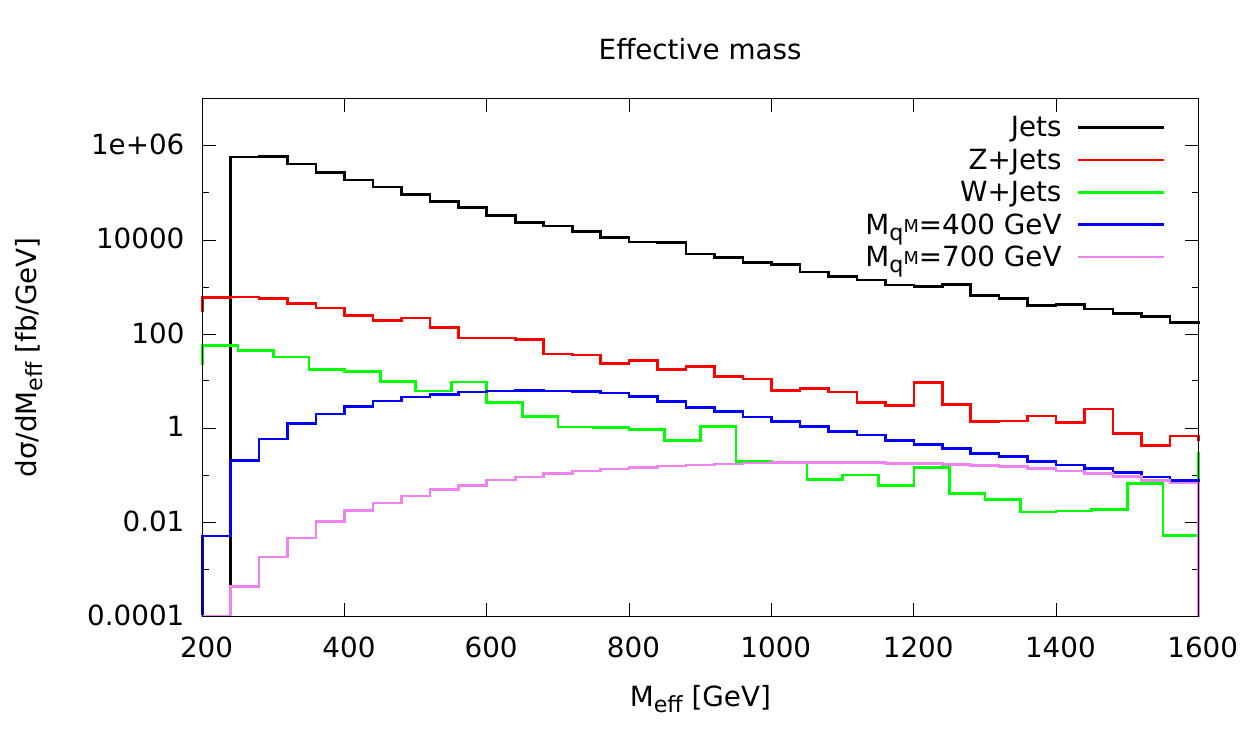}
	\caption{Missing transverse momentum (left panel) and effective mass (right panel) distributions for signal ($M=400$ and $700$ GeV) and background (jets, $Z$+jets, $W$+jets).}
\label{pt_mis}
\end{figure}

At this stage, we are equipped enough to develop a systematic
methodology of suppressing the SM backgrounds without drastically
reducing  the signal.  A fruitful perusal of such a
methodology requires that we carefully examine and compare the phase
space distributions of different kinematic variables for signal as
well as backgrounds discussed above. However, before we embark on the
mission to suppress the aforementioned backgrounds, it is important to
list a set of basic requirements for jets to be visible at
the detector.  It should be noted that any realistic detector
has only a finite resolution; this applies to both energy/transverse
momentum measurements as well as the determination of the angle of
motion. For our purpose, the latter effect can be safely
neglected
and we
simulate the former by smearing the energy with Gaussian
functions: 
\begin{figure}
\label{dphi}
	\centering
	\includegraphics[width=0.45\textwidth, angle=0]{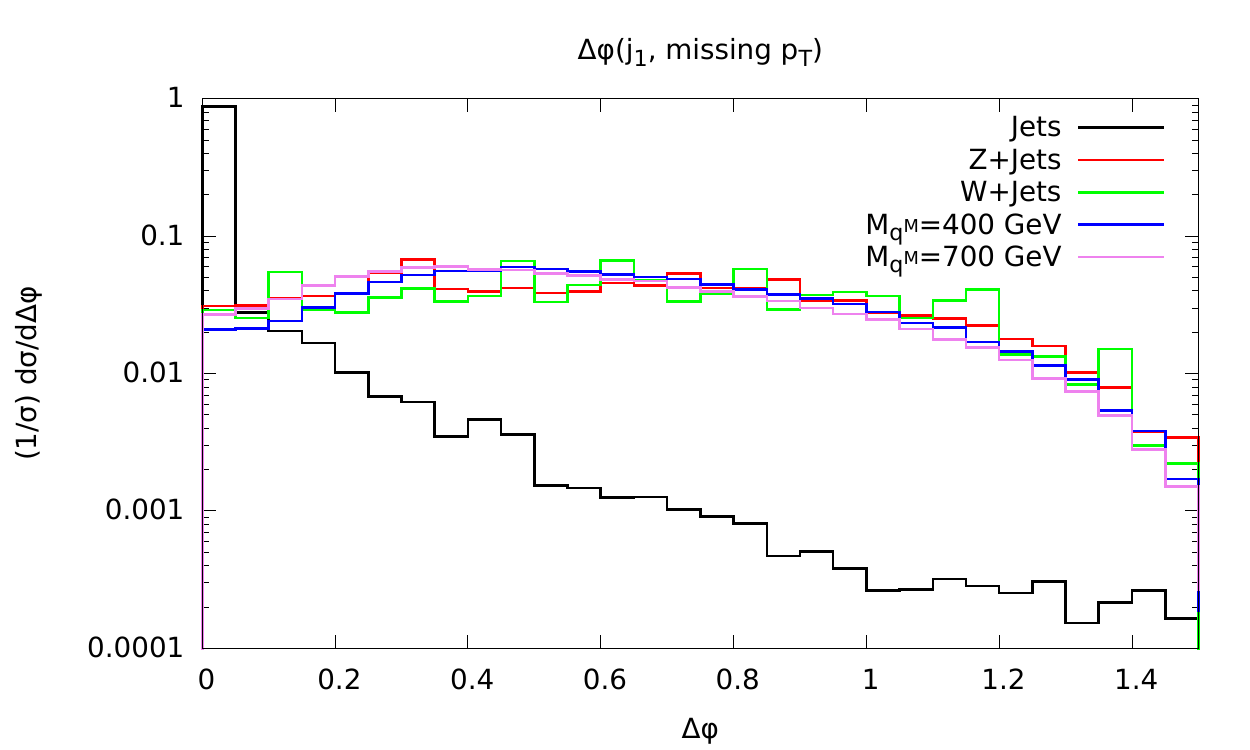}
	\includegraphics[width=0.45\textwidth, angle=0]{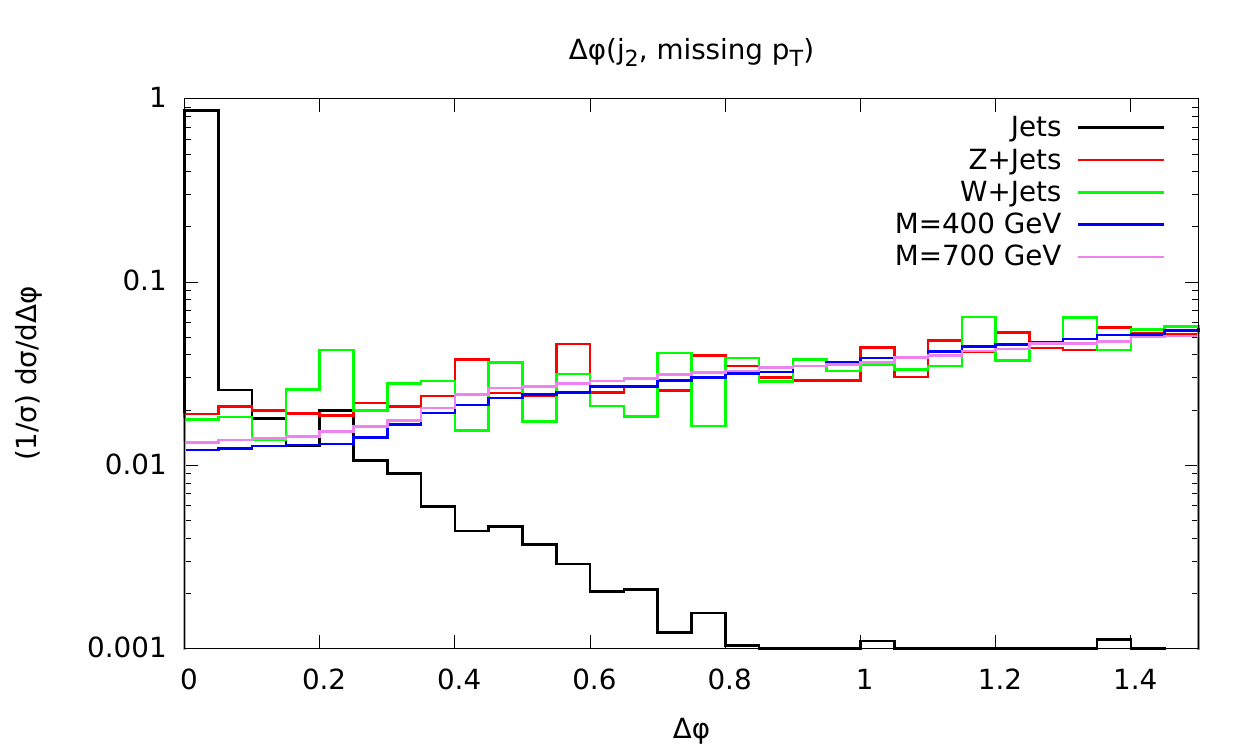}
	\caption{Normalized $\Delta \phi(jet,\vec p_T\!\!\!\!\!\!/~)$ distribution for signal and background.}
\end{figure}
\be
\frac{\Delta E}{E}=\frac{a}{\sqrt {E/{\rm GeV}}}\oplus b,
\ee
where,  $ a=100\%, b=5\%$, and $\oplus $ denotes a sum in quadrature.\cite{gsmear}
Keeping in mind the LHC environment as well as the detector configurations, 
we demand that, to be visible, a jet must have an 
adequately large transverse momentum and they are well inside 
the rapidity coverage of the detector, 
namely,
\be
p_T^{j} > 40~GeV \ ,
\label{cut:pT}
\ee
\be
|\eta_{j}| \leq 2.5 \ .
\label{cut:eta}
\ee
We demand that jets be well separated so that they can be identified as individual entities. To this end, we use the well-known cone algorithm defined in terms of a cone angle $\Delta R_{ij} \equiv \sqrt{\left (\Delta \phi_{ij}\right)^2 
+ \left(\Delta \eta_{ij}\right)^2} $, with 
$\Delta \phi $ and $ \Delta \eta $ being the azimuthal angular 
separation and rapidity difference between two particles.
Quantitatively, we impose
\be
\Delta R_{j \, j} > 0.7.
\label{cut:jj-iso}
\ee
Furthermore, the event must be characterized by a 
minimum missing transverse momentum defined in terms of the 
total visible momentum, namely,
\be
\not p_T \equiv \sqrt{ \bigg(\sum_{\rm vis.} p_x \bigg)^2 
                 + \bigg(\sum_{\rm vis.} p_y \bigg)^2 } \; > 20 \gev \ .
\label{cut:pt-mis}
\ee 
It has been discussed already that for some of the SM 
backgrounds, the hard (parton level) process does not even 
have a source of missing energy.  However,
the multi-jets final state could potentially be associated 
with a missing transverse momentum only on account of mis-measurements of the jets energies. A minimum requirement of the missing transverse momentum 
keeps these backgrounds well under control. 
The requirements summarized  in 
Eqs. (\ref{cut:pT}--\ref{cut:pt-mis}) constitute our 
{\em acceptance cuts}.

\begin{figure}
	\centering
	\includegraphics[width=0.7\textwidth, angle=0]{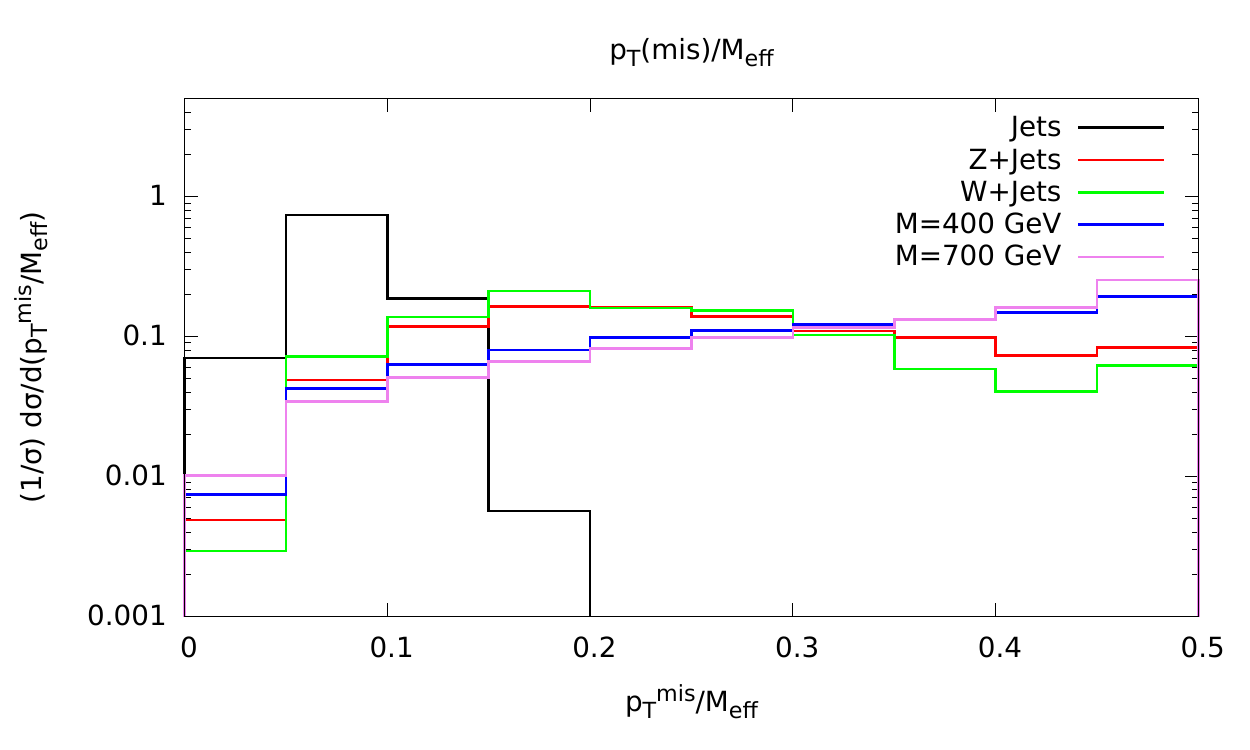}
	\caption{Normalized $p_T\!\!\!\!\!\!/~/M_{eff}$ distribution for signal and background.}
\label{ptmeff}
\end{figure}

With the set of acceptance cuts and detector resolution defined 
in the previous paragraph, we 
compute the signal and background cross-sections 
at the LHC  operating with $\sqrt{s} = 13 $ TeV 
respectively and display them in Table~\ref{tab:cross}.
Clearly, the backgrounds
are very large compared to the signal. The dominant SM background
contribution arises from multi-jets.  In order to
enhance the signal to background ratio, we study distributions of
different kinematic observables.
\begin{table}[h]
\begin{center}
\begin{tabular}{|c|c||c|c|c||c|}
\hline \hline
\multicolumn{6}{|c|}{Cross-section in fb}\\
\hline\hline
\multicolumn{2}{|c||}{Signal}  & \multicolumn{4}{|c|}{Background}
\\\hline
\multicolumn{2}{|c||}{$M_{q^M}$ [GeV]} & Jets & $Z$+jets & $W$+jets & Total \\
400 & 700 &  &  & & \\\hline
3.1$\times 10^3$ & 165.3 & 1.01$\times 10^8$ & 1.56$\times 10^5$ & 1.05$\times 10^4$ & 1.01$\times 10^8$ \\\hline\hline 
\end{tabular}
\end{center}
\caption{Signal and SM background cross-sections (in fb) after the acceptance 
cuts. We have also estimated $t\bar t Z$ background using AlpGen generator. Assuming invisible decay of the $Z$-boson, $t\bar t Z$ cross-section is estimated to be about 100 fb at the LHC with 13 teV center of mass energy. The other background contributions are orders of magnitude larger than $t\bar t Z$ contribution. As a result, we have not included $t\bar t Z$ contribution in this Table.}  
\label{tab:cross}
\end{table}
In Fig.~\ref{pt_jet}, we display the $p_T$ distributions of the signal and background jets after ordering them according to their $p_T$ ($p_T^{j_1}>p_T^{j_2}$). The left panel corresponds to the hardest jet and the right panel corresponds to the second hardest jet. From the shape of the $p_T$ distributions in 
Fig.~\ref{pt_jet} it is obvious that any harder $p_T$ cut on jets will simultaneously reduce signal as well as background. In Fig.~\ref{pt_mis}, we display the missing transverse momentum distribution (left panel) and effective mass distribution (right panel) for the signal and background. The effective mass is defined as the scalar sum of the transverse momenta of all the visible particles (in this case all jets with $p_T>40$ GeV), as well as the total missing transverse momentum, it can be expressed, in our case, through
\be
M_{\rm eff}=\sum_{j}p_T^{j}+p_T\!\!\!\!\!\!/~ \ .
\label{meff}
\ee
The missing transverse momentum distribution in the left panel of Fig.~\ref{pt_mis} shows that the multi-jets background is peaked at a relatively low $p_T\!\!\!\!\!\!/~$. Since, for this process, a missing transverse momentum can arise only from mis-measurement, this contribution can be suppressed significantly by introducing  a harder  $p_T\!\!\!\!\!\!/~$ cut. Whereas,  Fig.~\ref{pt_mis} (right panel) shows that harder $M_{eff}$ cut helps to reduce $Z/W$+jets background. In Fig.~\ref{dphi}, we present the normalized azimuthal angular distribution between the hardest-jet and $\vec p_T\!\!\!\!\!\!/~$ ($\Delta \phi(j_1,\vec p_T\!\!\!\!\!\!/~):$ left panel) and second hardest-jet and $\vec p_T\!\!\!\!\!\!/~$ (($\Delta \phi(j_2,\vec p_T\!\!\!\!\!\!/~):$ right panel). Fig.~\ref{dphi} shows that a lower bound on $\Delta \phi(j_1,\vec p_T\!\!\!\!\!\!/~)$ and $\Delta \phi(j_2,\vec p_T\!\!\!\!\!\!/~)$ clearly reduce multi jets background. Finally, we consider the ratio $ p_T\!\!\!\!\!\!/~/M_{eff}$, and in Fig.~\ref{ptmeff}, present the distributions in the same. The background peaks around $ p_T\!\!\!\!\!\!/~ /M_{\rm eff} \sim 0.1$ and it is obvious that it would be reduced significantly 
if a lower bound on this ratio is imposed t. In view of the characteristic distributions presented in Figs.~\ref{pt_jet} to \ref{ptmeff}, we summarized our final event selection criteria in Table~\ref{selection}.
\begin{table}[h]
\begin{center}
\begin{tabular}{||c||c|c||}
\hline \hline
Variable & Lower bound & Upper bound\\
\hline\hline
$p_T\!\!\!\!\!\!/~$ & 160 GeV & \\
$p_T(j_1)$ & 130 GeV & \\
$p_T(j_2)$ & 60 GeV & \\
$\eta_{j}$ & -2.5 & 2.5 \\
$Delta R_{j_1 j_2}$ & 0.7 & \\ 
$\Delta \phi(j,\vec p_T\!\!\!\!\!\!/~)$ & 0.4 & \\
$M_{eff}$ & 1000 GeV & \\
$p_T\!\!\!\!\!\!/~ / M_{eff}$ & 0.2 & \\\hline\hline
\end{tabular}
\end{center}
\caption{Selection cuts.}  
\label{selection}
\end{table}
In Table~\ref{cross_sec}, we summarize the signal and the SM background cross-sections after the imposition the selection cuts listed in Table~\ref{selection}. 
\begin{table}[h]
\begin{center}
\begin{tabular}{|c|c||c|c||c|}
\hline \hline
\multicolumn{5}{|c|}{Cross-section in fb}\\
\hline\hline
\multicolumn{2}{|c||}{Signal}  & \multicolumn{3}{|c|}{Background}
\\\hline
\multicolumn{2}{|c||}{$M_{q^M}$ [GeV]} & Jets & $Z/W$+jets  & Total \\
400 & 700 &  &  & \\\hline
111.6 & 51.8 & - & 400 & 400 \\\hline\hline 
\end{tabular}
\end{center}
\caption{Signal and SM background cross-sections (in fb) after the selection cuts.}  
\label{cross_sec}
\end{table}

\begin{figure}
	\centering
	\includegraphics[width=0.7\textwidth, angle=0]{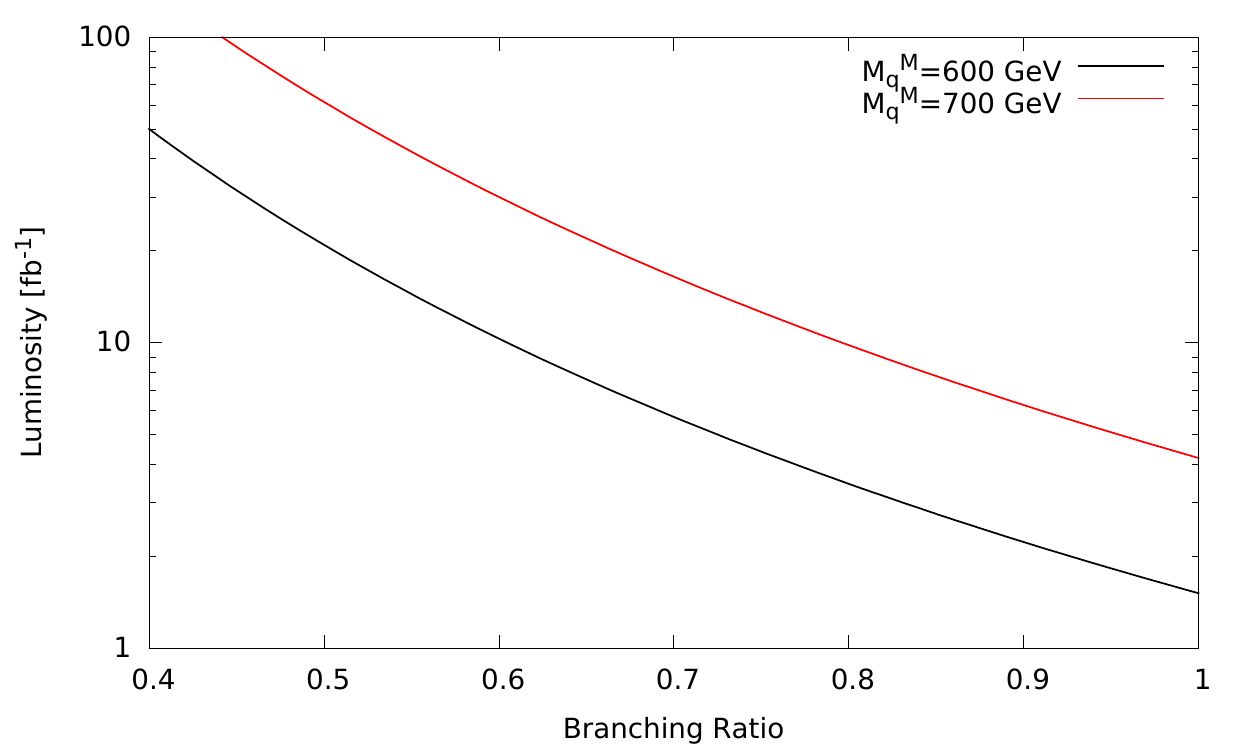}
	\caption{Required integrated luminosity for 5$\sigma$ discovery at the LHC with 13 TeV center-of-mass energy as a function of the branching ratio $ (q^M \to q \phi_S)$. for two different values of mirror quark masses (400 and 700 GeV).
\label{lumi400}}
\end{figure}

In order to calculate the discovery reach of the LHC with 13 TeV center of mass energy, we define the signal to be observable for a integrated luminosity ${\cal L}$ if,
\begin{itemize}
\item
\begin{equation}
\frac{N_{S}}{\sqrt{N_B+N_S}} \ge 5 ~~~~~~ {\rm for}~~~~~ 0< N_B \le 5 N_S,
\end{equation}
where, $N_{S(B)}=\sigma_{S(B)} {\cal L}$, is the number of signal (background) events for an integrated 
luminosity ${\cal L}$.
\item For zero number of background event, the signal is observable if there are at 
least five signal events. 
\item In order to establish the discovery of a small signal (which could be 
statistically significant i.e. $N_S/\sqrt{N_B}\ge 5$) on top of a large background, we need to know the 
background with exquisite precision. However, such precise determination of the SM background is beyond 
the scope of this present article. Therefore, we impose the requirement $N_B\le 5 N_S$ to avoid such 
possibilities.
\end{itemize}

\begin{figure}
	\centering
	\includegraphics[width=0.7\textwidth, angle=0]{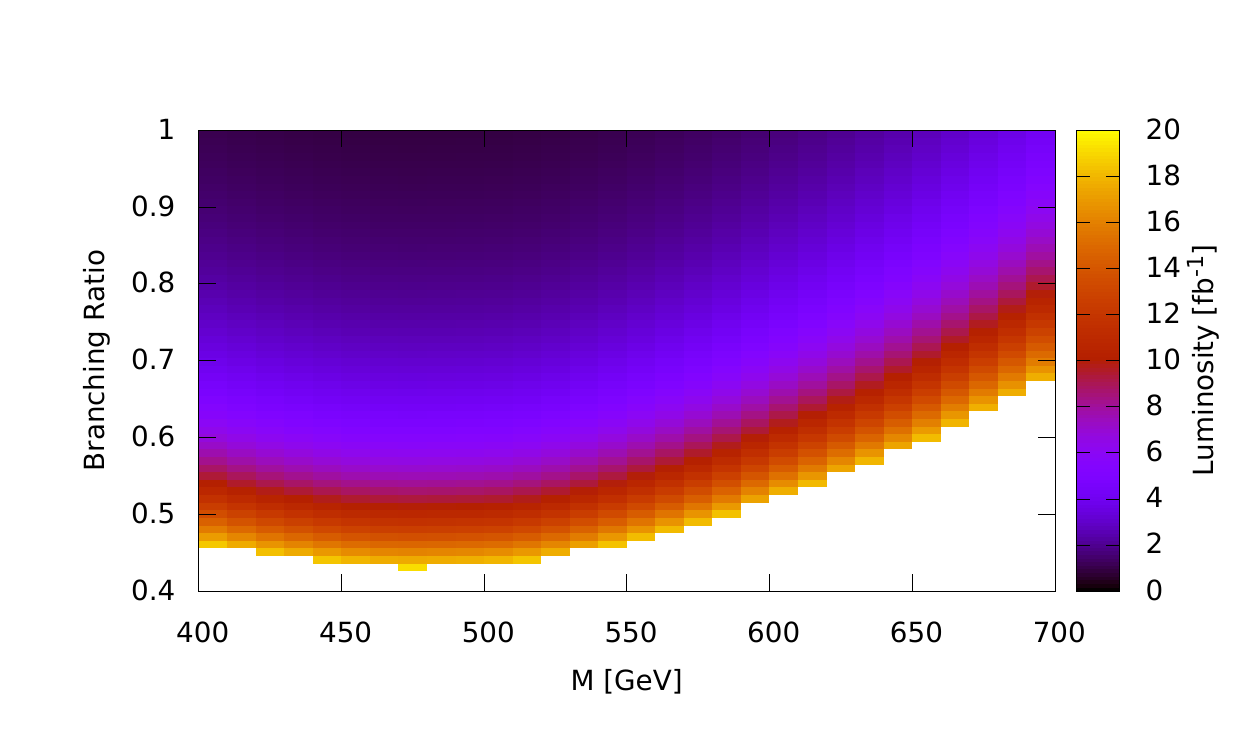}
	\caption{Required integrated luminosity (colour gradient) for 5$\sigma$ discovery at the LHC with 13 TeV center-of-mass energy is presented as a function of mirror-quark mass (along $x$-axis) and  the branching ratio $(q^M \to q \phi_S)$ (along $y$-axis).}
\label{lumi}
\end{figure}

The branching ratio of $q^M \to q \phi_s$ is a free parameter in this model. Therefore, in Fig.~\ref{lumi400}, we have presented required integrated luminosity for the discovery of mirror quark as a function of  the branching ratio $ (q^M \to q \phi_S)$ for two different values of mirror quark mass. In Fig.~\ref{lumi}, required integrated luminosity (color gradient) for 5$\sigma$ discovery at the LHC with 13 TeV center-of-mass energy is presented as a function of both mirror-quark mass (along $x$-axis) and  branching ratio $ (q^M \to q \phi_S)$ (along $y$-axis).

 {In this section, we have discussed the production and signature of the lightest mirror quark. This analysis is based on the assumption of prompt decays of the mirror quarks. In the framework of this model, the lightest mirror quark decays into a SM quark and $\phi_S$ with a Yukawa coupling which is a free parameter. Small values of this Yukawa coupling result in small decay widths and hence, a long lifetime for the lightest mirror quark. In this case, the produced mirror quarks hadronize before they decay. Depending on the smallness of the Yukawa coupling, the hadronized mirror quarks decay a few mm to a few cm away from the hard scattering point. The decay of a mirror quark gives rise to a missing particle and a jet. However, if the decays take place at a point different from the primary vertex then present LHC jet reconstruction algorithms categorize the resulting jets as "Fake Jets" \cite{fake1,fake2}. The LHC jet reconstruction algorithm is designed to distinguish jets produced in proton-proton collisions and  "Fake jets" not originating from hard scattering events. "Fake jets" come from different sources like the collision of one proton of the beam with the residual gas within the beam pipe, beam-halo events, cosmic rays e.t.c. The LHC jet reconstruction algorithm employs criteria like  the distribution of energy deposits by the jet, the shower shape and its direction, in particular the pointing to the interaction point to discriminate collision jets and "Fake jets". In the context of our model, if the mirror quark decays away from the interaction point then the resulting jets will not point towards the interaction point and hence, will be considered as "Fake jets" by the present LHC jet reconstruction algorithm. Moreover, in the absence of any information about the other decay product namely, the invisible scalar $\phi_S$, it will be very challenging to reconstruct the secondary vertex. Therefore, in order to study such events, a new algorithm for the jet reconstruction is required. It is beyond the scope of our parton level Monte Carlo analysis to study such events.}

\section{Conclusion}
 We have explored the new physics possibilities at the 13 TeV LHC run. The model used is well motivated, and was proposed to obtain tiny neutrino masses via the seesaw mechanism with the RH neutrino at the EW scale. The gauge symmetry is $SU(2)_w  \times U(1)_Y$; but the fermion as well as the Higgs sector is extended. For the fermions, we have both the left and right handed doublets, as well singlets under the $SU(2)_W$ gauge symmetry. In this model the RH quarks/lepton doublets, and the left handed singlets are called mirror quarks and leptons. The scalar sector of the original EW-scale $\nu_R$ model contains 2 triplets, one doublet and one singlet. However, the extended  EW-scale $\nu_R$ model contains: 2 triplets, 2 doublets (one coupled to the SM fermions and the other to mirror fermions), 4 singlets to accommodate the 125-GeV scalar and to discuss leptonic mixings.
 The model is derived using the above gauge symmetry, additional  global symmetries and discrete $A_4$ symmetry. As shown in the previous works, the model satisfies the EW precision constraints as well as the constraints from the 125 GeV Higgs data.

In this work, we have explored the implications of the model for the 8 and 13 TeV LHC. Since the model has colored quarks of chirality opposite to the SM quarks (the mirror quarks) and there is only one symmetry breaking scale, the usual EW scale $\sim$ 250 GeV, which gives masses to these mirror quarks, they cannot be heavier than $\sim$ 900 GeV from the unitarity of the Yukawa couplings. Thus these mirror quarks will be copiously produced at the LHC. Once produced, in the model, they can decay to ordinary quarks and singlet Higgs. At the interaction point if the corresponding Yukawa coupling is not tiny, the final states are two ordinary quarks, or two $b$ quarks and large missing energy  due to the escaping singlet Higgs. We have calculated the production cross sections time the branching ratios to ordinary light quarks or the b quarks. The relative branching ratios to the ordinary light quarks or the $b$ quarks are unknown in the model. So we have calculated the cross section times branching ratios as a function of the mirror quark masses for several values of the branching ratios. Comparing these with the corresponding experimental limit plots produced by the CMS Collaboration at the 8 TeV LHC, we find that the mass of the lightest mirror quark as low as $\sim$  {600} GeV is allowed  {for Br($q_M\to q \phi_S$) = 100\%. However, if the branching ratio of $q_M\to q \phi_S$ is 50\% or less then there is no bound from the LHC 8 TeV data.} 
{ Furthermore, this bound applies 
only to prompt decays of mirror quarks. For displaced decays with decay lengths greater than 1 mm or so, this bound is no longer valid and the mirror quark mass can be lower.}  {Assuming  prompt decays of mirror quarks,} we have also calculated the signal as well as the SM background at the 13 TeV LHC.  We find that the signal for the final state signature with a 5 sigma confidence level can be observed for the lightest mirror quark mass of $\sim$ 700 GeV with an integrated luminosity $\sim ~100 fb^{-1}$. 

The model has another interesting feature. For very tiny Yukawa couplings, the decays of these mirror quarks can produce displaced vertices with decay length in the range of cm or larger. These events characteristics are very different from the displaced vertices produced by b quarks for which average decay length is $\sim$ 0.5 mm.  Such unusual events may have been thrown out in the usual experimental analyses.  A suitable algorithm may need to be developed to look for such events.

\newpage 

\section*{Acknowledgments}
We are grateful to J. Haley, A. Khanov of the ATLAS Collaboration and G. Landsberg of the CMS Collaboration for several valuable
discussions. The work of SC, KG and SN was supported in part by  the US DOE Grant Number
DE-SC0010108. VH and PQH were supported in part by the
US DOE grant DE-FG02-97ER41027 and in part (PQH) by the Pirrung Foundation. SC is supported by the Pirrung Foundation.


\begin{thebibliography}{99}
\bibitem{Weinberg}
S.~Weinberg,
  Phys.\ Rev.\ Lett.\  {\bf 43}, 1566 (1979).

\bibitem{leeyang}
   T.~D.~Lee and C.~-N.~Yang,
  Phys.\ Rev.\  {\bf 104}, 254 (1956).
\bibitem{pqnur}
P.~Q.~Hung,
	Phys.\ Lett.\ B {\bf 649}, 275 (2007)
	[hep-ph/0612004].
\bibitem{LR}
  J.~C.~Pati and A.~Salam,
  Phys.\ Rev.\ D {\bf 10}, 275 (1974)
  [Phys.\ Rev.\ D {\bf 11}, 703 (1975)];
  R.~N.~Mohapatra and J.~C.~Pati,
  Phys.\ Rev.\ D {\bf 11}, 2558 (1975);
G.~Senjanovic and R.~N.~Mohapatra,
  Phys.\ Rev.\ D {\bf 12}, 1502 (1975);
  G.~Senjanovic,
  Nucl.\ Phys.\ B {\bf 153}, 334 (1979);
  S.~Chakdar, K.~Ghosh, S.~Nandi and S.~K.~Rai,
  Phys.\ Rev.\ D {\bf 88}, no. 9, 095005 (2013).  	
  \bibitem{WR}
  V.~Khachatryan {\it et al.} [CMS Collaboration],
  Eur.\ Phys.\ J.\ C {\bf 74}, no. 11, 3149 (2014)
  [arXiv:1407.3683 [hep-ex]].
  

\bibitem{hung2}
 V.~Hoang, P.~Q.~Hung and A.~S.~Kamat,
  Nucl.\ Phys.\ B {\bf 877}, 190 (2013)
  [arXiv:1303.0428 [hep-ph]].

\bibitem{hung3}
V.~Hoang, P.~Q.~Hung and A.~S.~Kamat,
  Nucl. Phys. B {\bf 896} (2015) 611-656,
  [arXiv:1412.0343 [hep-ph]].
\bibitem{pqtrinh}
  P.~Q.~Hung and T.~Le,
  JHEP {\bf 1509}, 001 (2015)
  [arXiv:1501.02538 [hep-ph]].
  \bibitem{pqaranda}
  	A.~Aranda, J.~Hernandez-Sanchez and P.~Q.~Hung,
	JHEP {\bf 0811}, 092 (2008)
	[arXiv:0809.2791 [hep-ph]].	 
 \bibitem{fake1} 
  [ATLAS Collaboration],
  ATLAS-CONF-2012-020, ATLAS-COM-CONF-2012-018.
\bibitem{fake2} 
  G.~Aad {\it et al.} [ATLAS Collaboration],
  JINST {\bf 8}, P07004 (2013)
  [arXiv:1303.0223 [hep-ex]].
 

  
\bibitem{montvay}
  I.~Montvay,
  DESY-87-147;
  Phys.\ Lett.\ B {\bf 199}, 89 (1987);
  DESY-87-077.
  
  \bibitem{hung4}
  P.~Q.~Hung, T.~Le, V.~Q.~Tran and T.~C.~Yuan,
  JHEP {\bf 1512}, 169 (2015)
  doi:10.1007/JHEP12(2015)169
  [arXiv:1508.07016 [hep-ph]].  
  
\bibitem{h_ww_122013}
	S.~Chatrchyan {\it et al.}  [CMS Collaboration],
	JHEP {\bf 1401}, 096 (2014)
	[arXiv:1312.1129 [hep-ex]].
	
\bibitem{h_zz_4l_122013}
  S.~Chatrchyan {\it et al.} [CMS Collaboration],
  Phys.\ Rev.\ D {\bf 89}, no. 9, 092007 (2014)
  [arXiv:1312.5353 [hep-ex]].
	
\bibitem{h_bb_102013}
	S.~Chatrchyan {\it et al.}  [CMS Collaboration],
	Phys.\ Rev.\ D {\bf 89}, 012003 (2014)
	[arXiv:1310.3687 [hep-ex]].
	
\bibitem{h_tautau_012014}
  S.~Chatrchyan {\it et al.} [CMS Collaboration],
  JHEP {\bf 1405}, 104 (2014)
  [arXiv:1401.5041 [hep-ex]].
\bibitem{ATLAS_jets} 
  G.~Aad {\it et al.} [ATLAS Collaboration],
  JHEP {\bf 1409}, 176 (2014)
  [arXiv:1405.7875 [hep-ex]].
\bibitem{CMS_jets} 
  S.~Chatrchyan {\it et al.} [CMS Collaboration],
  Eur.\ Phys.\ J.\ C {\bf 73}, no. 9, 2568 (2013)
  [arXiv:1303.2985 [hep-ex]].
\bibitem{Khachatryan:2015wza} 
  V.~Khachatryan {\it et al.} [CMS Collaboration],
  JHEP {\bf 1506}, 116 (2015)
  [arXiv:1503.08037 [hep-ex]].
\bibitem{Mangano:2002ea}
  M.~L.~Mangano, M.~Moretti, F.~Piccinini, R.~Pittau and A.~D.~Polosa,
  JHEP {\bf 0307}, 001 (2003).
 \bibitem{gsmear}
 G.~Aad {\it et al.} [ATLAS Collaboration],
  arXiv:0901.0512 [hep-ex],
  G.~L.~Bayatian {\it et al.} [CMS Collaboration],
  J.\ Phys.\ G {\bf 34}, 995 (2007).
  
 
 
 
 
 
 

\end{thebibliography}
\end{document}